# New experiments on the study of the electric activity of He II at the excitation of second-sound waves

A.S. Rybalko, V.A. Tikhiy, A.S. Neoneta, K.R. Zhekov

*B.I. Verkin Physico-Technical Institute of Low Temperatures of the NAS of Ukraine*

*47, Lenina Prosp., Kharkov, 61103, Ukraine*

E-mail: rybalko@ilt.kharkov.ua

**Abstract.**

We present the results of new experiments on the observation of the electric response of superfluid helium (He II) at the excitation of second-sound waves in it. We used an acoustic resonator with dielectric case, in which the measurements of the electric and thermal signals were carried on simultaneously due to the disposition of a measuring electrode and a bolometer on the face wall. We measured the potential differences between the face electrodes, as well as between ring metallic electrodes disposed along the resonator case which were insulated from one another and from helium. It is shown that the distribution of the electric potential along the resonator coincides with that of the temperature in a standing second-sound wave. It is established that the ratio $\Delta T/\Delta U$ in a second-sound wave is a constant equal to $2.3 \cdot 10^4$ K/V in the whole regions of temperatures, frequencies, and heat flow powers under study.

## 1. Introduction

At present, our knowledge of superfluid He II is based on the Landau phenomenological two-fluid theory and the Onsager--Feynman rule of the macroscopic quantization of momenta. According to the theory, the motions of the normal and superfluid parts of the fluid are independent at temperatures below T=2.17 K. However, the nature of the internal non-phenomenological mechanism of superfluid motion is not yet clear and requires new experimental facts. The superfluid motion by itself is a consequence of processes running on the microscopic level that result in the appearance of a motive force. The vectors of microscopic forces are added and form a macroscopic force. The force is a result of the action of fields. The task of experiments is to establish the nature of microscopic forces in liquid helium

Namely the theoretically predicted property of independence of the normal and superfluid motions can be used to clarify the physics of fields. The essence of experiments must consist in the artificial creation of a relative flow of the components in the fluid bulk surrounded by gages for the registration of a field. Second sound as a specific quantum-mechanical phenomenon consisting in weakly decaying oscillations of the temperature and the entropy in superfluid helium is most suitable for this purpose. A second-sound wave in He II realizes the case where the total flow of the momenta of the superfluid and normal components is zero, $\rho_s\mathbf{V_s}+\rho_n\mathbf{V_n}=0$, i.e., the momentum flows of two components are. Here, $\rho_s$ and $\rho_n$ are, respectively, the densities of the superfluid and normal components, and $\mathbf{V_s}$ and $\mathbf{V_n}$ are their velocities. The above-presented idea was used in [1], where the experimental of the second sound in superfluid helium was undertaken to elucidate whether the electric field arises inder conditions of the opposite flows $\rho_s$ and $\rho_n$.

The measuring set-up includes a second-sound resonator combined with an electric condenser. Plates of the condenser were joined with the input of a voltmeter, and no external electric field was supplied to them. At the excitation of a second-sound wave, charges were induced on the condenser plates. The experiments showed that the amplitude of oscillations of the potential difference were proportional to the amplitude of oscillations of the temperature in a second-sound wave and were equal to $\Delta U \sim 10^{-7}$ V by the order of magnitude. At the same time, the value of $\Delta U$ was independent of the material of measuring electrodes and exceeded the values of thermo-e.m.f. characteristic of the metals used as electrodes by several orders [2-3].

The results in [1] could be explained, by assuming the presence of a permanent electric moment in the system. But, according to the modern ideas, a helium atom, being a spherically symmetric system, has no electric moment. We cannot compare the results obtained in [1] with those of other experimental works, since the "pyroelectric" properties of liquid helium were not studied earlier. However, the exceptionality of the electric properties of He II would be indicated by works, in which the influence of an electric field on the flow velocity of helium through narrow slits [4] and on the transfer on a film [5] was observed. In addition, a lot of works dealt with a high-precision measurement of the polarizability of liquid helium are known (see, e.g., [6-7] and references therein). Experiments on the measurement of the dielectric constant $\varepsilon$ might indicate a deviation from the Clausius–Mossotti law, if such a moment were present in a system. As known, some effective $\varepsilon$ consisting on the orientational and polarization parts of the polarizability was measured in condensed systems including electric moments. By the Langevin--Debye model, the orientational part inversely proportional to the temperature is equal to the field strength multiplied by the dipole moment ($\sim pE/T$), and the polarization part is proportional to the square of the field strength ($\sim \alpha E^2/2$). Here, $\alpha$ is the polarizability of a helium atom. In this case, the results of measurements of $\varepsilon$ turn out to be dependent on the measuring voltage (field) and the dipole moment p [8-9]. The analysis of many works implies that, only in [7], the data on the polarizability referred to the same density above and below the $\lambda$-point show a deviation from the Clausius–Mossotti law. It is worth noting that if the induction signal of 100-200 nV observed in [1] is considered as the reaction to the formation of a domain with a vary small orientational part $pE/3kT \ll \alpha E^2/2$ ($U = E_{coord} \cdot L \ll 1$ μV), then the electric voltages used in the measurements of $\varepsilon$ in [6] and other analogous works are more higher than the induced signals. Here, L is the resonator length. For example, at a voltage of 16 V and a relative accuracy of measurements equal to $10^{-7}$ [6], the potential of ~0.1 μV induced by the orientational part can remain unnoticed despite the high accuracy of measurements. Unfortunately, the measuring field strength was not indicated in [7]. Thus, we have no unique answer concerning the existence o fthe moment. However, the problem of search for moments in condensed helium is urgent, since it is known that the temperature dependence of the heat capacity has λ-shape for the majority of substances, where the second-order phase transitions are observed, and a transition itself is related to the presence of some moments. The general regularities of the heat capacity, permittivity, permeability, and the effect of the measuring field on ε and μ in the substances, where the second-order phase transitions are observed, were considered in [8-9].

Since a helium atom has no electric dipole moment, the electric response arising at the thermal motion in a superfluid fluid turned out unexpected. Experiments [1] initiated the appearance of theoretical works [10-29], where several models of the observed phenomenon of the “electric activity” of He II were developed. However, no satisfactory explanation of the experimental was obtained: the values of induced signals in experiments [1] were by several orders higher.

The present work, being a repetition and a continuation of the experiment in [1], is devoted to a more detailed study of conditions for the appearance of the electric response and to its specific features. The thorough analysis and the wide discussion (see [10-29]) of work [1] and the subsequent works [30-31] revealed a number of problems and some sources of possible errors, so that the effect cannot be considered unambiguous without additional experimental verification. The most significant problem consists in the correct measurement of from the source with high internal resistance signals in the nanovolt interval. The subsequent problems are the capacitive interferences from the alternating electric potentials of a heater and a bolometer. Since the electric signal was weak, it would seem that the effect of electric activity of superfluid helium is absent. Therefore, in order to obtain the reliable absolute values of ΔU, we need the careful calibration of gages and amplifying channels, which requires the design and the fabrication of a special facility. In addition, we use a new construction of the second-sound resonator allowing us to simultaneously register oscillations of the temperature and of the electric potential. One more distinction is related to the ability to measure the spatial distribution of the electric response in a second-sound wave and to register the phase shift between oscillations of the temperature and the electric potential polarity.

## 2. Experimental method.

***Cooler.*** Experiments were carried out on an evaporative cooler in the temperature interval 1.3 ÷ 2.3 K. The evaporation chamber was continuously filled with liquid helium through a throttle. The productivity of an evaporative cooler was at least 5 mW at a temperature of 1.3 K. The evaporation chamber and the working bath, to which pure helium was condensed from a separate balloon, were fabricated of a single piece of copper. The working chamber contained a second-sound resonator and a resistance thermometer submerged in liquid helium.

***Second-sound resonator***. We use a resonance method of registration of the second-sound. Due to a high quality factor of the modes (Q=200-4000), this method allows us to accumulate a high energy density W in the resonator on resonance frequencies at the release of insignificant powers by a heater w: W=Qw [32].

In Fig. 1, we present a scheme of the second-sound resonator and the basic diagram of measurements allowing us to simultaneously register oscillations of the temperature and the electric potential. The resonator is cylinder 6 of 25 mm in length, whose external and internal diameters are 30 and 4.9 mm, respectively. The length equal to 25 mm was chosen on the basis of a reasoning that the wave profile is distorted on a path of only several cm due to the strong dependence of the second-sound on its amplitude [33]. The resonator cylinder was produced of Stycast and closed by face lids made of brass. In the cylinder body symmetrically to its axis, four identical cylindical electrodes 7 (abcd) are placed. They are insulated from one another and have an internal diameter of 6 mm. Thus, these electrodes were separated from the fluid by a plastic layer 0.5 mm in thickness and cannot be directly heated by a second-sound wave.

To excite second-sound waves, we used low-inertia thin-film heater 5. It was sprayed on a ceramic support, fixed by Stycast in a hollow of one of the lids, and feeded from generator 1. The heater resistance $R_h$=(5041±20) Ohm below 2.3 K in the temperature interval under study. Generator 2 was used for calibration of amplifying channels.

In the holow of the second lid, we glued thin-film thermometer-bolometer 8 registering oscillations of the temperature $\Delta T$ and electrode 9, being a plate of the measuring condenser. As its second plate, we took the lid with a heater or one of electrodes abcd. In the last case, this allowed us to register the spatial distribution of the electric response $\Delta U$ along the whole length of the resonator, by connecting the required electrodes in their turn.

In the mounting of the resonator, a thin layer of vacuum lubricant served as an impermeable element between the case and one of the lids, and the second lid was simply pressed. Without a lubricant, a lot of splits were present between a lid and the case, through which helium filled in the resonator very rapidly. The experiments showed that, at such a means of the connection with the external bath, the quality factor of the given resonator decreases rapidly, as the heater power increases, and does not exceed 500-700 even in the region 1.95-2.05 K, where the absorption of the second sound is minimum. The presumably strong motion of the normal and superfluid components in the split under the action of thermal flows in the direction perpendicular to the acoustic wave is a reason for the decrease of the quality factor. This reasoning is supported by the fact that the quality factor stops to be dependent on the power in a wider range. Moreover, ΔT was greater and dT/dw was steeper, if we introduced paper spacer 13 between a lid and the case, through which the resonator 0.5 $cm^3$ in volume was filled with a fluid for 30-40 min at 1.4-1.6 K. The use of filling supersplit 13 and weakly heat-conducting plastic walls reduces the energy losses in a thermal wave due to the connection with the bath to a minimum. The high quality factor of the resonator (up to Q≤4000) was ensured by the parallelism of the faces and the polishing of the internal cylindrical wall.

Thus, we measured oscillations of the temperature on a face wall of the resonator and the potential difference between the faces of the resonator, between a face and one of the cylindrical electrodes, or between two cylindrical electrodes.

***Receivers of the second sound.*** Both receivers 8 and 9 were selected from a pilot batch (SMD resistor) of resistance thermometers which are a $RuO_2$ film 1×3 $mm^2$ in size applied on a ceramic support and protected by a lac layer. Lac was dissolved in acetone, and the gages were placed in a 15-% HCl solution for some time in order that the resistance at 4.2 K was 4-6 kOhm. At liquid-helium temperatures, the resistance of $RuO_2$ depends strongly on the temperature, which allowed us to use this gage as a rapidly operating thermometer in the measurement of oscillations of the temperature in a second-sound wave. Preliminarily, we measured the temperature dependence of the resistance of both gages 8 and 9. In the interval of temperatures under study, their resistance varied from 5 to 40 kOhm. The sensitivity of the thermometers was $1/R\cdot(dR/dT)=1.72\text{-}6.70\ K^{-1}$ in the temperature interval 1.4- 2.3 K, which allowed us to surely the amplitude of oscillations of the temperature, ΔT, in a second-sound wave down to $10^{-6}$ K.

***Receivers of induced charges in a second-sound wave*** were the condensers with capacitance C. One of their plates was insulted electrode 9. (In other experiments, film electrode 9 can be also used as a thermometer.) As the second plate of a condenser, we took the heater lid or one of electrodes abcd. As was shown in [1], a charge q=ne=CΔU was induced on the plates of a condenser at the propagation of the second sound, where ΔU is the induced potential difference. In this case, the effect did not depend on the material of plates (gold, copper, brass, and ruthenium oxide) and was much greater than that in metals [2].

***Amplifying channels.*** Amplitudes of oscillations of the temperature and the electric response were amplified by two independent channels composed of pre-amplifiers 11 and synchronous detectors 12 (a Lock-in 5204 and a Lock-in nanovoltmeter 232V). (In Fig. 1, we show only one amplifying channel.) We also used a selective nanovoltmeter 233. Generator 1 supplied an alternating current with variable f with a minimum step of 0.001 Hz to heater 5. Since the heating occurs during the positive and negative half-periods, the second-sound waves were excited, as usual, at the double frequency 2f. Respectively, the signal from the frequency doubler (it is not shown in the figure) was supplied on the base legs of amplifiers. A part of experiments was carried out with the polarization of a heater with direct current. In this case, the frequency doubler was not applied.

The manifestation of “the electric activity” of superfluid helium in a second-sound wave indicates, possibly, the appearance of a cophasal displacement of the charges of atoms in the fluid, which is a source of the induction of charges on the condenser plates. By the preliminary evaluations based on the measurements in [1], this source has a high internal resistance (~$10^9$ Ohm), which is a substantial physical characteristic of the observed effect. At such high resistance, the input capacitance equal to several picofarads forms a low-frequency filter with descent starting from several Hz. In addition, a finite resistance of the insulation in a connecting cable can easily deteriorate the performance of an amplifier with superlow current of the input signal due to leakages. The same is true for the thermometric channel. Both problems are solved by means of the use of a protecting electrode and the pre-amplifiers designed by the impedance transformation principle [33]. In other words, the line (cable) connecting electrode 9 with the pre-amplifier input has double screen 10 (see Fig. 1). The internal screen is connected with the output of a follower, which is the first block of the pre-amplifier. This excludes efficiently the currents of resistive and capacitive leakages, if we realize conditions, under which the potential difference between the signal wire and its environment is equal to zero. To attain these conditions, the subsequent cascades of a pre-amplifier are embraced by a positive feedback and, in essence, are the active filter. To decrease the noise, we fed the pre-amplifiers from a separate power source. The pre-amplifiers have the amplification coefficient ≤10 and low input currents and well suppress the cophasal signals. The positive feedback and the active compensation ensure the linearity of the transmission band up to 2.5 kHz at a level of 3 dB.

The use of generator 2 allows us “to compensate” the input capacitance [33], to measure the amplification coefficient and the transmission band, and to evaluate the values of signals and noises relative to the calibration voltage directly in the experiment in the nanovolt interval [34].

***Compensation of the input capacitance.*** The process of compensation consists in the selection of feedback parameters that should be such that the capacitance of the intermediate screen is charged through the feedback up to the same charge at the same phase, as the measuring capacitance of the resonator [33]. To this end, we supply the voltage $\Delta U_{in}^{кал}$=100 nV from calibration generator 2 via chopper 3 and attenuator 4 to the resistance R=50 Ohm (Fig. 1) and tune the feedback. The resistance $R_{in}$ plays the role of the internal resistance for the electric activity source. Under the compensated capacitance of a cable, the signal has a maximum value, which is fixed by the amplifier as $(\Delta U_{in}^{кал})_1$. This potential difference corresponds to the induced charge q=C $(\Delta U_{in}^{кал})_1$ on the plates of the measuring condenser C.

As was indicated above, to measure the density of induced charges, it is necessary to know the value of measuring capacitance. Therefore, after the process of compensation of the cable capacitance, we measured C. To make this, we connect the calibration capacitance $C_{cal}$ with the input of the pre-amplifier. In this case, the amplitude of a signal decreases to the value $(\Delta U_{in}^{cal})_2$. The measuring capacitance C was calculated from the charge conservation law

$$C\,(\Delta U_{in}^{cal})_1=(C+C_{cal})\,(\Delta U_{in}^{cal})_2\,, \qquad (1)$$

where indices 1 and 2 denote the measured values of signals before and after the connecting of the calibration capacitance. It was established that such a means allows one to decrease the input capacitance down to 0.1 pF. Knowledge of the value of capacitance C gives a possibility to correctly evaluate the induced charge.

We emphasize that, within the used scheme, we can determine the polarity of the electric response relative to oscillations of the temperature in a second-sound wave. The arrival of a wave and the heating of a bolometer were accompanied by a decrease of its resistance and, hence, the voltage on it, whose polarity is set by the polarity of the feeding battery of a bolometer. As the feeding current changes the direction, the phase of a variable voltage on a bolometer varies by 180 degrees. To measure the polarity of the electric response during the experiment, it was sufficient to compare the phases of a signal from thermometer 8 and a signal from electrode 9. To make this, we supply the sinusoial signal from thermometer 8 amplified to 0.3 V to the base input of a Lock-in, where it was compared with the signal of the electric response of electrode 9. Thus, we can measure the polarity and observe directly the connection between oscillations of the temperature and oscillations of the electric response by the amplitude and by the phase.

***Calibration of amplifying channels.*** We performed the calibration and the verification of the amplification coefficients $K_{ampl}$ of both amplifying channels before and during the measurements. Knowledge of $K_{ampl}$ is necessary for the exact determination of the absolute values of oscillations of the temperature and the electric response. In Fig. 1,b, we present the record of a graduated variation of the voltage of a calibration generator under conditions, when the friendly signal was absent. In this case, the voltage from generator 2 was supplied to the base input of an amplifier. A voltage of 5 nV corresponds to the output voltage equal to 0.036 V.

In the course of the experiment, the frequency of one of the generators was scanned, whereas another generator had fixed values of frequency and amplitude $U_{in}^{cal}$. If the frequency of the calibration generator falls in the sensitivity band of a Lock-in, we observe beats with amplitude $U_{out}^{cal}$ on the output (see the inset in Fig. 1,c). The amplitude of beats is maximum at the point, where If the frequency of the calibration generator coincides with the known frequency of the scanned generator feeding a heater. In Fig. 1,c, we show the record of arising beats: the sum of the calibration voltage and the signal of generator 1 simulating a high-resistance source. Beats against the background of a signal of the electric response will be shown below at the discussion of results. The amplification coefficient of the whole channel is determined as the ratio of the voltage of beats on the output and the calibration voltage on the input:

$$K_{ampl} = U_{out}^{cal} / U_{in}^{cal} \qquad (2)$$

Thus, the true value of $\Delta U_{true} = \Delta U_{èçì}$ out/K$_{ampl}$ for a signal is calculated from the measured value of $\Delta U_{meas}$ with regard for the amplification coefficient K$_{ampl}$.

***Analysis of noises and interferences.*** The value of noises was estimated by means of the comparison of the noise track with the calibration signal (the insert in Fig. 1,b and 1,c). In the insert in Fig. 1,c, we present the record in the 1-Hz band for a graduated change of the calibration signal in the channel of the electric response at various input signals from 0 to 15 nV. As is seen, the noise track gives ~3-4 nV in the 1-Hz band and does not depend on the input signal, which determines the limiting sensitivity of amplifying systems to the voltage.

It follows from Fig. 1 that a capacitive interference between heater 5 and gage 9 can be a source of false signals and interferences. Despite the fact that the pre-amplifiers suppress efficiently the cophasal interference, the ground wire of a heater passed along the intermediate screen and was connected with the ground near the generator. Analogously, the ground wire of a bolometer passed along the intermediate screen and was connected with the ground near the pre-amplifier. After the mounting of the resonator in the cryostat, we determined the maximum admissible values of voltages on a heater and a bolometer, at which the interference becomes to exceed the intrinsic noises of the amplifier.

After the compensation of the capacitance, the measurements under conditions when a signal with a frequency of 2f is supplied to the base leg of the synchronous detector showed that the interference does arise and becomes comparable with with the noise track at the peak voltage on a heater exceeding 4 V. In this case, the interference is independent of the frequency. But if the heater is polarized by a constant voltage, on which a variable voltage with frequency f is imposed, then the interference becomes comparable with the noise at a variable voltage of ~0.3 V. For this reason, the voltage supplied to a heater in real experiments with liquid helium did not exceed these limiting values in all the cases.

At the simultaneous measurement of oscillations of the temperature and the electric response, the interference can appear due to the capacitive connection between gage 8 and electric response gage 9. The coupled (simultaneous) measurements showed that such an interference becomes comparable with the noise track, when oscillations of the voltage on gage 8 due to received second-sound waves exceed ~30 μV. Therefore, the measuring current of thermometer 8 was selected so that the value of $\Delta U^T$ on it was less than 30 μV in the whole ranges of temperatures and voltages supplied to a heater. The additional test for the capacitive connection between gages 8 and 9 was the switching-off of the feeding current of a bolometer: signal of oscillations of the voltage on a bolometer disappeared, whereas the electric response did not vary.

The above-described measures of calibration and caution allowed us to reliably measure the absolute values of variable voltage in the nanovolt interval from sources with high internal resistance.

## 3. Resonances of oscillations of the temperature and the electric induction in a second-sound wave.

We illustrate the process of preparation and the course of the experiment in Fig. 2. As the temperature decreases from 4.2 K to 1.3 K, the working chamber of the resonator begins to be filled slowly by superfluid helium through a superslit. A heater was feeded by a variable voltage with amplitude $U_h$, whose frequency was scanned in the required range (Fig. 2,d). Simultaneously, we recordered the potential difference (the signal from the amplifying channel $U_{out}^E$ , Fig. 2,b) and oscillations of the temperature $U_{out}^T$ (Fig. 2,c). If necessary, we calibrated the amplifying channels. For example, the input calibration signal of 14 nV became $U_{out}^{cal}$=0.1 V on the output of the amplifier of the electric response (Fig. 2,b), which gave us the value of amplification coefficient. After the filling of the resonator by superfluid helium, the signal related to the second-sound appeared. Simultaneously, a potential difference (in the given case, between electrode 9 and the heater lid) arose. The clear resonances of oscillations of the temperature and the potential difference on electrode 9 relative to the ground were observed. In order to be sure that the electric signal on electrode 9 is not a result of the capacitive interference, we switched-off the feeding of bolometer 8 for some time. The variable voltage on the bolometer disappeared, whereas the electric signal on electrode 9 traced, as before, the amplitude-frequency characteristic of the second-sound mode. The change of the feeding polarity of a bolometer had also no effect on the electric signal. Thus, we established that the electric signal is related to thermal flows in a second-sound wave, rather than to a capacitive interference. The suspicion concerning the thermo-e.m.f. of the metals of electrodes was also true as well. For example, the thermo-e.m.f. is equal to 0.07 μV/K for gold at 4.2 K and decreases sharply with the температурой [3]. In the experiments in [1] and in the present work, the maximum amplitude of oscillations of the temperature is ~3 mK, which would give the electric signal equal to $0.07\cdot 0.003=2.1\cdot 10^{-4}$ μV.

Second sound was excited in the frequency range from 250 to 2000 Hz, which corresponds to the frequency of generator 1 from 125 to 1000 Hz. In this frequency range, we observed several even and odd harmonics of the second sound. The presence of harmonics, the influence of the temperature on their position, and a number of peculiarities of the observation are presented in Fig. 3 for oscillations of the temperature and for oscillations of the electric potential. We registered the high-quality resonance peaks of oscillations of the temperature $\Delta U^T_i$ of the first five modes. At the measurement of the potential difference $\Delta U^E_i$ between electrode 9 and the heater lid, the signals were observed only on odd harmonics – the first, third, and fifth ones. The amplitudes of modes of the temperature signal decrease, as the harmonic number increases. Proportionally, the electric signals on odd modes decreased so that the ratio $\Delta U^T_{1,3,5,}/\Delta U^E_{1,3,5}$=const was the same for the first, third, and fifth resonances. The first we could assume is that the equalities $\Delta U^T_1/\Delta U^E_1$= $\Delta U^T_3/\Delta U^E_3$= $\Delta U^T_5/\Delta U^E_5$=const are a result of the capacitive interference. But if we are faced with the interference or the thermo-e.m.f., the absence of signals on even modes is not understandable. In the measurements of the potential difference between electrodes 9 and c, we observed signals on all modes, but the ratios $\Delta U^T_{1,2,3,4,5,}/\Delta U^E_{1,2,3,4,5}$ were not identical.

The position of resonance frequencies depends on the temperature, and the calculation of the propagation velocity of signals $\Delta U^T$ and $\Delta U^E$ by the formula V=2Lf/n gives values exactly coinciding with the known values of second-sound velocity [32, 35-36]. Here, L is the resonator length, f is the frequency of a harmonic, and n is the number of a harmonic. In Fig. 3,b, we present the temperature dependences of signals at a constant power of the heater. Near points, we indicated the resonance frequencies. As is seen, the amplitudes of signals $\Delta U^T$ and $\Delta U^E$ decrease, as the temperature increases. At every separate temperature, the proportionality $\Delta U^T_1/\Delta U^E_1$= $\Delta U^T_3/\Delta U^E_3$= $\Delta U^T_5/\Delta U^E_5$ held true, but the constant was different for different temperatures. This fact proves that we deal with a new phenomenon, rather than with some interference.

Above the λ-point, we observed only noises, which constitute up to ~6 μK by ΔT and up to ~4 nV by ΔU. It is characteristic that noises constitute ~1 μK by ΔT below 2.17 K.

In Fig. 4,a and 4,c, we give the amplitude-frequency characteristics ΔT and ΔU for the first mode at several powers of the heater. In Fig. 4,b and 4,d, the same amplitudes as functions of the power at the resonance frequency ar epresented for different conditions of the connection of the resonator with the external bath: through the superslits formed by paper spacer 13 between the face and the case (curves 1) and through 10-μm slits (curves 2).

We observe typical resonance curves, whose amplitude increases with the power and are well fitted by a Gauss function. The upper and lower curves in Fig. 4,c are obtained at the heater power 3.22 mW/cm$^2$. They differ only in that the lower curve was measured, when the input of a pre-amplifier was connected with a capacitance of 10 pF, which is equivalent to the resistance $1/\omega C$= 40 MOhm. As is seen, this yields the strong diminution of a signal, i.e., the source of signals has a high internal resistance.

A single curve of the electric response represents the 10-nV calibration signals at frequencies of 392.4 and 396.9 Hz. The amplitudes of signals $\Delta T$ and $\Delta U$ increase with the power w supplied to the heater by a practically linear law (Fig. 4,a-c, data 1). The fact that the electric signal turned out proportional to $w \sim U^2_h/R_h$, rather than to $U_h$, proves additionally the absence of a capacitive interference from the heater. We note that Figs. 2 and 3 represent the primary measured data, whereas Fig. 4,b and 4,d show the true values of oscillations of the temperature and the electric response obtained with regard for the amplification coefficient $K_{ampl}$ measured experimentally (see Section 2).

The linear mode of the amplitude of oscillations of the temperature as a function of the power for the connection of the resonator with the bath through a supersplit was observed up to 3.5 mK at 1.3 K and up to 2 mK at 2 K. The comparison with the experimental data of other researchers indicates that the amplitudes of oscillations of the second sound (their absolute values $\Delta T$) obtained in the present work were higher at the corresponding powers of thermal flows and increased with the мощностью more rapidly than those in [32, 35-36]. Possibly, this is related to the circumstance that the quality factors of resonators in [32, 35-36] were less. Unfortunately, the values of quality factors were not presented in those works; or if even the quality factors were given, the absolute values of $\Delta T$ were absent. The second reason can consist in the lower losses through the lateral wall of the resonator in our experiment due to its less area and greater thickness.

The measurement of the quality factor Q at resonances indicates that it depends strongly on the temperature: Q~200 at T~1.3 K and increases to ~3500-3700 at T~2.0 K. While approaching the $\lambda$-point, we observe again a sharp drop in the quality factor. Such a behavior agrees with the known temperature dependence of the absorption of the second sound presented in [37] on the basis of plenty experimental data.

However, both the linear mode of the amplitude of oscillations of the temperature and the quality factor significantly depend on the connection of the resonator with the bath through a system for the filling with liquid helium. The quality factor of modes was ~211 (Fig. 4,a and 4,c), which corresponds to the accumulated energy W=wQ=0.68 W/cm$^2$. This value is lower than the threshold of excitation of the Kolmogorov spectrum by almost one order [38]. In the case where the paper microporous spacer was lacking and the macroscopic through holes (10-20 μm) were present between the face and the case of the resonator, the quality factor decreased sharply, as the power increased. The corresponding dependences $\Delta T$(w) and $\Delta U$ (w) at a resonance frequency are shown by curves 2 (Fig. 4,b and 4,d). These dependences reveal the characteristic peculiarities at w>1.5 mW/cm$^2$ which were observed also in [32] at somewhat higher powers. First, we observe a deviation of ΔT and ΔU from a linear law. But then, at some value of w, the dependence becomes again linear. The approximation of these data to ΔT→0 and ΔU→0 gives the same intercept on the x-coordinate for the thermal flow density. In [32], it was assumed that this fact can be related to the processes of formation of eddies arising in the presence o fthe counterflow of the superfluid and normal components through slits. The reasson for such an effect remains unclear. We note that, at the measurement of ΔT in this case, we observe a hysteresis, as w increases or decreases. But no hysteresis was observed at the measurement of ΔU.

Curves 1 and 2 in Fig. 4 show that the change of conditions of the experiment (the replacement of a superslit, through which of the resonator is filled, by an ordinary slit) change basically the values of ΔT in a second-sound wave. In this case, the potential difference is also changed. This fact allows us to assert that the observed potential difference is related to the wave process of second sound and is not an artefact.

By the two-fluid theory, the amplitude of oscillations of the temperature in a running second-sound wave at the thermal flow density W is

$$\Delta T = W/\rho C V_2 \,. \qquad (3)$$

Here, ρ is the helium density, C is its heat capacity, and $V_2$ is the second-sound velocity. In the standing second-sound wave, this value is greater due to the quality factor of the mode [32]:

$$\Delta T = wQ/\rho C V_2 \,, \qquad (4)$$

where w is the thermal flow density created by a heater per unit area of the resonator cross-section.

The comparison of experimental data on ΔT for resonators with different systems of connection with the external bath and the estimates of oscillations of the temperature in a second-sound wave by formula (4) indicate that the results given by curve 1 (Fig. 4,b) are in good agreement with the theory for a resonator with superslit. Our data on $\Delta T(Q\cdot w)_{res}$ also well agree with the available data on $\Delta T(W)_{pulse}$ [39-40] obtained by the pulse method for running waves.

As for the manifestation of the unusual electric properties of superfluid helium, we have shown the appearance of a source of “electric activity” with high internal resistance below the λ-point under the creation of the relative motion of the components in fluid helium. By measuring the potential difference between different electrodes of the resonator, we have established that the polarity of $\Delta U^E$ and the direction of the temperature gradient in a second-sound wave are interconnected. When the temperature of thermometer 8 (the velocity vector of the normal component is directed to the receivers) is minimal, the potential of electrode 9 is positive relative to the ground. On the contrary, when the temperature of thermometer 8 is maximal, the negative charge arises on face electrode 9. The polarity is independent of the direction of a current in a bolometer. This fact proves additionally that the registered signal is not a capacitive interference.

### 4. Spatial distribution of the value of electric response.

Due to the presence of many electrodes, we were able to measure the potential difference between points of the resonator $\Delta U^E_{ik}$, whereas the amplitude of oscillations of the temperature was measured only at a single point on the face of the resonator $\Delta T_i$ (it corresponds to $\Delta U^T_i$ on a bolometer). The identical value of $\Delta U^T_i/\Delta U^E_{ik}$ for the first, third, and fifth modes (see Section 3) indicates the existence of a direct connection between the thermal and electric characteristics of a second-sound wave. For these odd modes, $\Delta T_i$ coincides with $\Delta T_{ik}$. Therefore, it seemed of interest to compare $\Delta T_{ik}/\Delta U^E_{ik}$ at different points in the bulk of the resonator in a standing second-sound wave. For this purpose, we measured the potential difference between one of the ring electrodes and electrode 9 or between two ring electrodes. The results were compared with the difference of the amplitudes of oscillations of the temperature in a second-sound wave between those points.

The results are presented in Fig. 5. At the first harmonic ($f_1$=392 Hz) (see Fig. 5,a), the maximum signal of the potential difference was observed between electrode 9 and the heater lid. A somewhat weaker signal was measured between electrode 9 and ring electrode a, and, while approaching electrode 9 the value of ΔU at this harmonic becomes less and less. Any capacitive interference on rind electrodes cannot lead to these results, since no signal was observed both above and below the λ-point outside of the resonance at the same voltages on a heater.

At the excitation of sound at the third-harmonic frequency $f_3$=1176 Hz (Fig. 5,c), when the resonator length was equal to three half-wavelengths, the maximum signal was registered between electrodes c and 9 and between electrodes a and 9. But if this operation is repeated at the frequency, when the resonator length was equal to five half-wavelengths ($f_5$=1960 Hz), then the maximum signal was registered between electrodes b and 9 and between electrodes d and 9. Thus, we observed the clear resonances of the potential difference ΔU on the ring electrodes in the presence of a second-sound wave, which confirms the presence of the effect over the whole bulk of the resonator.

So, we may conclude that, in a standing second-sound wave, the fluid is polarized in the bulk inhomogeneously. The potential difference was maximum on the electrodes, the difference between which constituted a half-wavelength. The experimental spatial distribution of the potential difference was compared with the appropriate spatial distribution of the temperature ΔT (x,t) calculated by the known formula for a second-sound wave [32]:

$$\Delta T\ (x,t)=\Delta T_0\cos[2\pi ft-2\pi x/\lambda-\alpha]. \qquad (5).$$

Here, x is the coordinate along the resonator, t is the time o fthe propagation of a wave from the source to a point x, f is the frequency of a second-sound wave, λ is the wavelength, α is the initial phase of a wave, and $\Delta T_0$ is the maximum amplitude of oscillations on a bolometer. For a standing wave, $\Delta T\ (0,\ t)=\Delta T_0$, and α is equal to zero or multiple to 2π. If $\Delta T_0$ is determined experimentally, we can calculate the amplitude of oscillations of the temperature $\Delta T_k$ at any point x of the resonator by formula (5), and the propagation time t can be determined by using the second-sound velocity. Thus, $\Delta T_{ik}= \Delta T_0- \Delta T_k$.

In Fig. 6, we show the results of comparison of the experimental values of $\Delta U_{ik}$ (points) with $\Delta T_{ik}$ (continuous curve) calculated by formula (5) for the first, second, and third harmonics. The values are well correlated with one another for all harmonics. The ratio $\Delta T_{ik}/\Delta U_{ik}$ between two points in the resonator bulk is the `same with an accuracy of 20% (see Section 5). Moreover, if $\Delta T_{ik}=0$ between two points along the sound excitation path, then $\Delta U_{ik}=0$, which is observed in the pair modes by measuring these quantities between the faces of the resonator. At the same time, the signal between electrodes c and 9 is nonzero in the second mode. This fact indicates uniquely the following: charges of the same sign are generated on the faces of the resonator.

The potential differences between the cylindrical electrodes, which were thermally insulated from the fluid by a plastic spacer, repeated the wave process of second sound. This confirms once more that the thermo-e.m.f. of metals is extraneous to our results. Thus, the comparison of the spatial difference ΔT with the potenrial difference ΔU allowed us to interpret the obtained data. We have established that oscillations of the potential difference between the electrodes follow oscillations of the temperature (entropy) in a standing second-sound wave in phase along the whole length of the resonator. Based on the analysis of the data on $\Delta U^T_{ik}/\Delta U^E_{ik}$ for various pairs of electrodes, we can assert that the observed electric signal is induced by the processes running in He II at the propagation of a heat wave, and it is not a consequence of some capacitive interference. The reliability and the soundness of our results are confirmed by the reproducibility of data on different resonators and under different conditions.

The measured values of potentials on the plates of a condenser, which arise on the opposite faces of the resonator at the excitation of the second sound, allow one to calculate the density of induced charges and to determine the strengths of electric fields and the internal wave resistance of the source of electric activity. These questions will be considered elsewhere.

## 5 **Connection between oscillations of the temperature and oscillations of the electric potential.**

On the basis of the data obtained, we can establish the connection of thermal and electric characteristics of He II. Since we measured oscillations of the temperature $\Delta T(I_i,T_i,f_i)$ in a second-sound wave and, respectively, oscillations of the induced voltage $\Delta U(I_i,T_i,f_i)$, it seems natural to construct the plot of ΔT as a function of ΔU.

In Fig. 7, we present such dependences under the identical conditions (heater current $I_i$,, bath temperature $T_i$, and second-sound frequency $f_i$) which were measured at different times. Figure 7,a shows the primary data of a single experiment: measured voltage on a bolometer $\Delta U^T$ as a function of the voltage on electrode 9, $\Delta U^E$, at different temperatures. As is seen, the proportionality holds between the voltages, but the slope angle depends on the temperature. But if we transform the voltage on a bolometer into the amplitude of oscillations of the temperature in a second-sound wave, then all data lie on the single curve (Fig. 7,b). This proves once more that we deal with the electric response of superfluid helium to a heat wave, rather than with some capacitive interference. This plot contains also the data (open symbols) on $\Delta U(I_i,T_i,f_i)$ taken from ring electrodes, for which $\Delta T(I_i,T_i,f_i)$ were calculated, but not measured. The amplitudes of oscillations of the temperature along the resonator opposite to the middle of each of the ring electrodes were taken from the rated data for the n-th harmonic (see Fig. 6, Section 4). As is seen, the dependence of ΔT on ΔU is practically linear. The difference in the slope angles between various data of the present work and work [1] is ~20 %, which is related, probably, to a specific feature of the construction of the resonator: the presence of a plastic wall, whose effect on the response signal is difficult to be considered.

The linearity of plots and the independence of the slope angle on the heater power, bath temperature, and frequency testify to the direct proportional connection between thermal and electric processes in a second-sound wave. It is natural to consider ΔT as the energy of a heat wave, $k\Delta T/2$, per one degree of freedom and the product of the charge e by ΔU as the energy of the electrostatic interaction responsible for the appearance of polarization charges.*** ***(This reasoning was kindly communicated to us by S. Fil'.) Under conditions of a periodic variation in the entropy, these both different types of energy are transformed into each other. The slope angle tangent $\Delta T/\Delta U$ possessing the dimension of the ratio of the elementary charge to the Boltzmann constant, K/V, turned out independent of the temperature in the interval 1.4 – 2.1 K and is equal to $2.3 \cdot 10^4$ K/V. We note that this value is close to $2e/k = 2.3188 \cdot 10^4$ K/V, where e is the electron charge, and k is the Boltzmann constant. In other words, the ratio of the amplitudes of oscillations of the temperature and the electric response (the potential difference) in a second-sound wave is equal to the ratio of the electron charge to the Boltzmann constant to within a constant which is an integer (~2) and independent of the temperature. The question about whether this result is accidental or has a profound physical sence requires the additional theoretical and experimental studies.

The experiments with second sound indicate that the He II system acquires the order characterized by the appearance of a macroscopic dipole moment proportional to the temperature gradient. The electric response registered on the plates of a condenser is caused by the motion of atomic charges, so that the electron subsystem shifts coherently relative to the nuclear subsystem. Hence, we may conclude that the internal degree of freedom of atoms reveals itself in a second-sound wave. It is worth noting that the calculation of the thermodynamic characteristics of fluids involves usually only the degrees of freedom related to their thermal motion and does not takes the “internal” degrees of freedom into account.

Helium atoms have no dipole moment and the lowest polarizability among the other atoms. However, when liquid helium transits into the superfluid state, the temperature gradient induces, as was shown by the performed experiments, the potential difference. The nature of internal electric fields arising in superfluid helium requires a special analysis and will be considered in a separate publication.

## Conclusion.

- We have confirmed the effect of electric activity of He II in a second-sound wave by means of the simultaneous measurement of oscillations of the temperature $\Delta T$ and the electric potential $\Delta U$;
- The careful calibration of gages and amplifying channels allows us to obtain the reliable absolute values of $\Delta T$ and $\Delta U$;
- It is established that the ratio $\Delta T/\Delta U$ is a constant equal to $2.3 \cdot 10^4$ K/V in the whole regions of temperatures, frequencies, and powers of a thermal flow for a second-sound wave under study;
- For the first time, a spatial distribution of the electric response in a standing second-sound wave is determined;
- The polarity of the induced potential relative to the gradient of temperatures is found.

We are grateful to S. Semenov for the fabrication of pre-amplifiers and to A. Andreev, A. Kovalev, A. Kuklov, V. Maidanov, L. Mel'nikovskii, V. Mineev, V. Natsik, A. Patashinskii, E. Pashitsky, Yu. Poluektov, E. Rudavskii, S. Ryabchenko, S. Sokolov, V. Sivokon', S. Shevchenko, G. Sheshin, S. Fil', and O. Usatenko for the discussion of results.

The work is supported by the grant UNTTs 5211.

References

1. A.S. Rybalko, *Fiz. Nizk. Temp*., **30**, 12, 994, (2004).
2. F.J. Blatt. Physics of electronic conduction in solids. 1968., F.J. Blatt. Physics of electronic conduction in solids. 1971 [in Russian] (fig. 80, p. 235, M:«Mir»).
3. Cusack N., Kendall P., Proc. Phys. Soc., **72**, 899 (1958).
4. W. Neidhardt, J. Fajans, Phys. Rev., vol **128**, 496,(1962)
5. B.N. Eselson at al. Phys Lett. **47A**. 29, (1974).
6. M. Chan, M. Ryschkewitsch, H. Mayer, JLTP, vol. **26**, (1977)
7. J. Stankowski, S. Sitarz, Z. Trybula, V. Kempinski, and T. Zuk, *Acta Phys. Polon* **A70**, 29, (1986).
8. B.A. Strukov, **Soros**. **Obraz**. **Zh**., **2**, 7, (1996).
9. S. A. Gridnev, **Soros**. **Obraz**. **Zh**., **6**, 8, (2000).
10. A.M. Kosevich, *Fiz. Nizk. Temp* **31**, 1, 50, (2005).
11. A.M. Kosevich, *Fiz. Nizk. Temp* **31**, 10, 1100, (2005).
12. L.A. Melnikovsky, Cond-mat/0505102 v1. 2.(2005)v.3(2006). L.A. Melnikovsky, J. Low. Temp. Phys. 148, 559, (2007).
13. V.D. Natsik, *Fiz. Nizk. Temp* **33**, 12, 1319 (2007), *Fiz. Nizk. Temp* **34**,7, 625, (2008).
14. V.D. Khodusov, Visnyk KhNU, **642**, 79 (2004*)*, Visnyk KhNU, **721**, 31, (2006).
15. E. A. Pashitsky, S.M. Ryabchenko, *Fiz. Nizk. Temp.* **33**, 12, (2007)
16. V.M. Loktev, M.D. Tomchenko, *Fiz. Nizk. Temp* **34**, 4-5, 337, (2008).
17. E.D. Gutliansky, *Fiz. Nizk. Temp* **35**, 10, 956, (2009).
18. S.I. Shevchenko, A. S. Rukin, JETP Letters **90**, 46, (2009), S.I. Shevchenko, A. S. Rukin, Fiz. Nizk. Temp. 36, 186, (2010).
19. V.P. Mineev, J Low Temp Phys **162,** 686–692, (2011).
20. V.D. Khodusov, A.S. Naumovets, Visnyk KhNU, **899**, #2/46/, 97, (2010).
21. V. M. Loktev and M. D. Tomchenko, Ukr. J. Phys. **55**, 901, (2010).
22. M. D. Tomchenko, J. Low Temp. Phys. **158**, 854, (2010).
23. S. I. Shevchenko and A. S. Rukin, Pis'ma Zh. Eksp. Teor. Fiz.[JETP Lett.] **90**, 46 (2009).
24. S. I. Shevchenko and A. S. Rukin, Fiz. Nizk. Temp. **36**, 186, (2010) [Low Temp. Phys. 36, 146 (2010)].
25. E. A. Pashitsky and A. A. Gurin, Zh. Eksp. Teor. Fiz. **138**, 1103, (2010) [JETP 111, 975 (2010)].
26. V. P. Mineev, J. Low Temp. Phys. **162**, 686, (2011).
27. V. M Loktev, M. D Tomchenko, J. Phys. B At. Mol. Opt. Phys. **44**, 035006, (2011).
28. M. D. Tomchenko, PHYSICAL REVIEW B **83**, 094512, (2011).
29. R.V. Grigorishin, B.I. Lev, Ukr. J. Phys. **53**, 7, 635, (2008).
30. A. Rybalko, E. Rudavskii, S. Rubets, V. Tikhiy, V. Derkach, S. Tarapov, J. Low. Temp. Phys. **148**, 527, (2007).

31. A.S. Rybalko, S.P. Rubets, **31**, 7, 820, (2005).
32. V.P. Peshkov, Zh. Teor. Eksp. Fiz., **18**, 950, (1948).
33. U. Tietze, Ch. Schenk. Poluprovodnikovaya schemotekhnika. [in Russian], Moscow: «Mir», (1998), p.466.
34. L.I. Slabkii, Methods and Devices for Limiting Measurements in Experimental Physics [in Russian], Moscow: «Nauka», (1973).
35. V.P. Peshkov, Zh. Teor. Eksp. Fiz., **16**, 1000, (1946). *Zh. Teor. Eksp. Fiz.,* **23**, 686, (1952).
36. K. Zinov'eva, Zh. Teor. Eksp. Fiz., **25**, 235, (1953).
37. Russell J. Donnelly, Physics Today **34**, October 2009.
38. A. N. Ganshin, V. B. Efimov, G.V. Kolmakov, L. P. Mezhov-Deglin, P.V. E. McClintock, PRL **101,** 065303, (2008)
**39.** P. Zhang, M. Murakami, Phys. Rev. B **74**, 024528, (2006)
40. Fuzier, S., Van Sciver, S.W., Cryogenics **44**, 211, (2004).

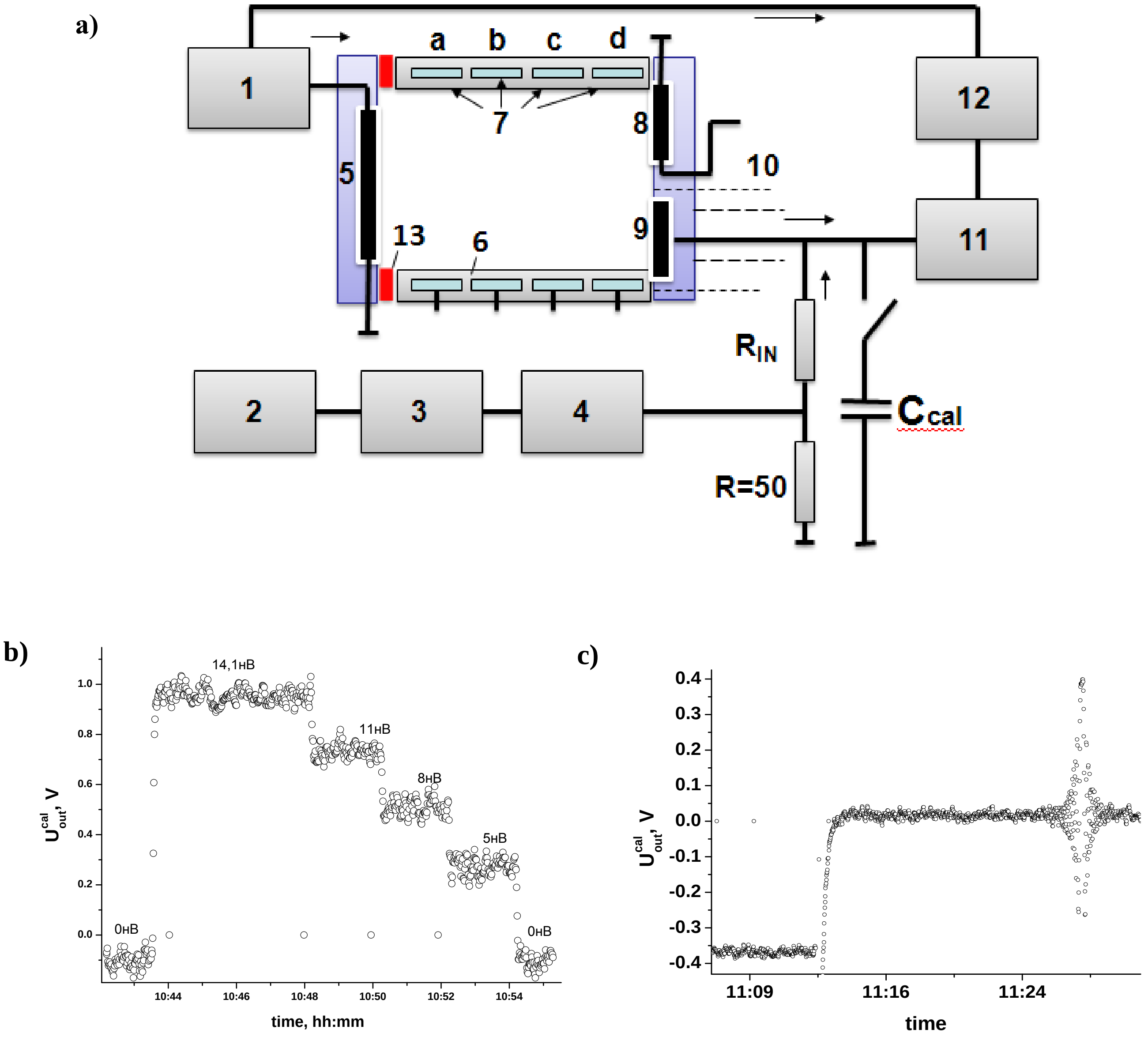

Fig.1 The scheme of experiment: 1 - scanning generator; 2 - calibrating generator; 3 - chopper; 4 - attenuator 120dB; 5 –heater; 6 –insulator; 7 – ring electrodes; 8 - thermometer; 9 –electrode of the measuring capacitor; 10 –double shield; 11 - preamplifier; 12 - Lock-in 5204; 13 –paper spacer. The amplifying tract of bolometer is not shown. a), b), c) – illustrate calibration of an amplifying path (the explanation in the text).

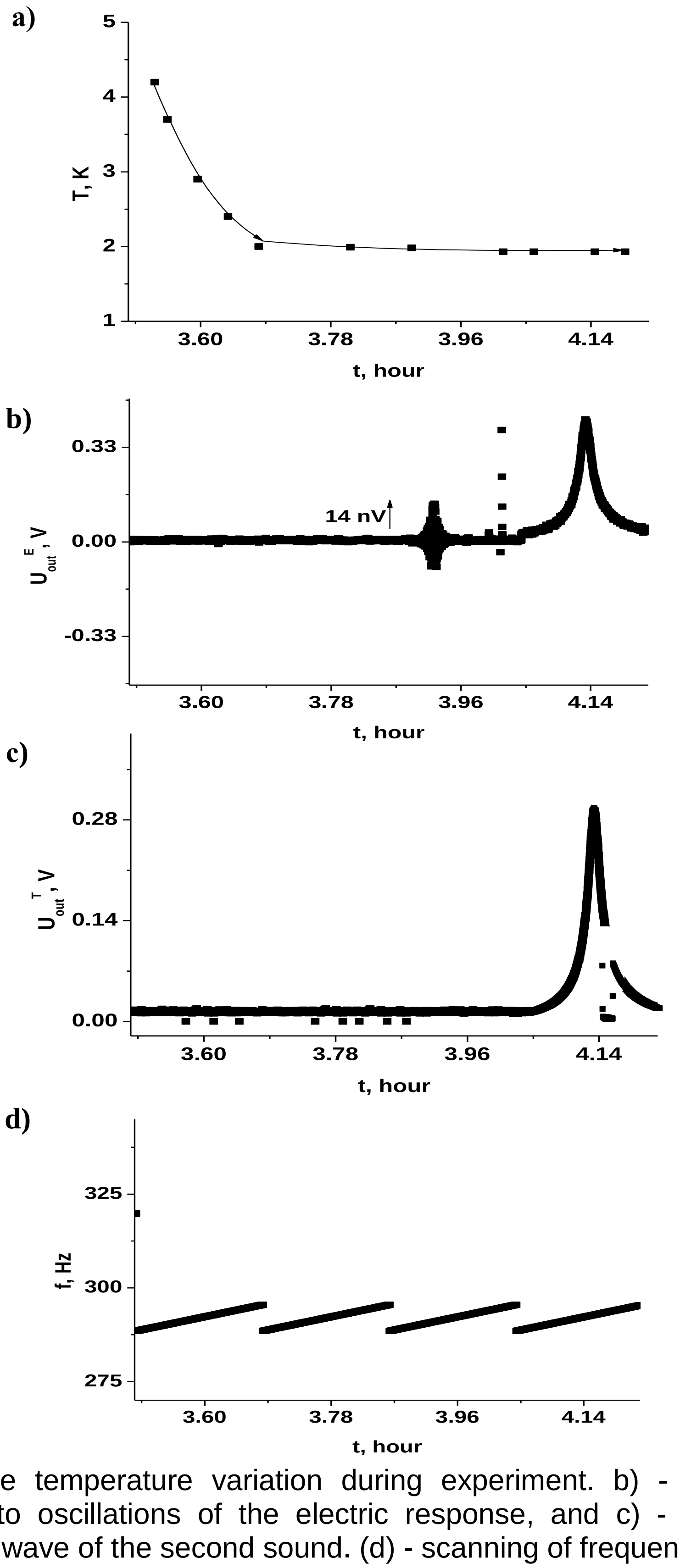

Fig.2. a) – The temperature variation during experiment. b) - the typical signals corresponding to oscillations of the electric response, and c) - to the temperature oscillations in a wave of the second sound. (d) - scanning of frequency.

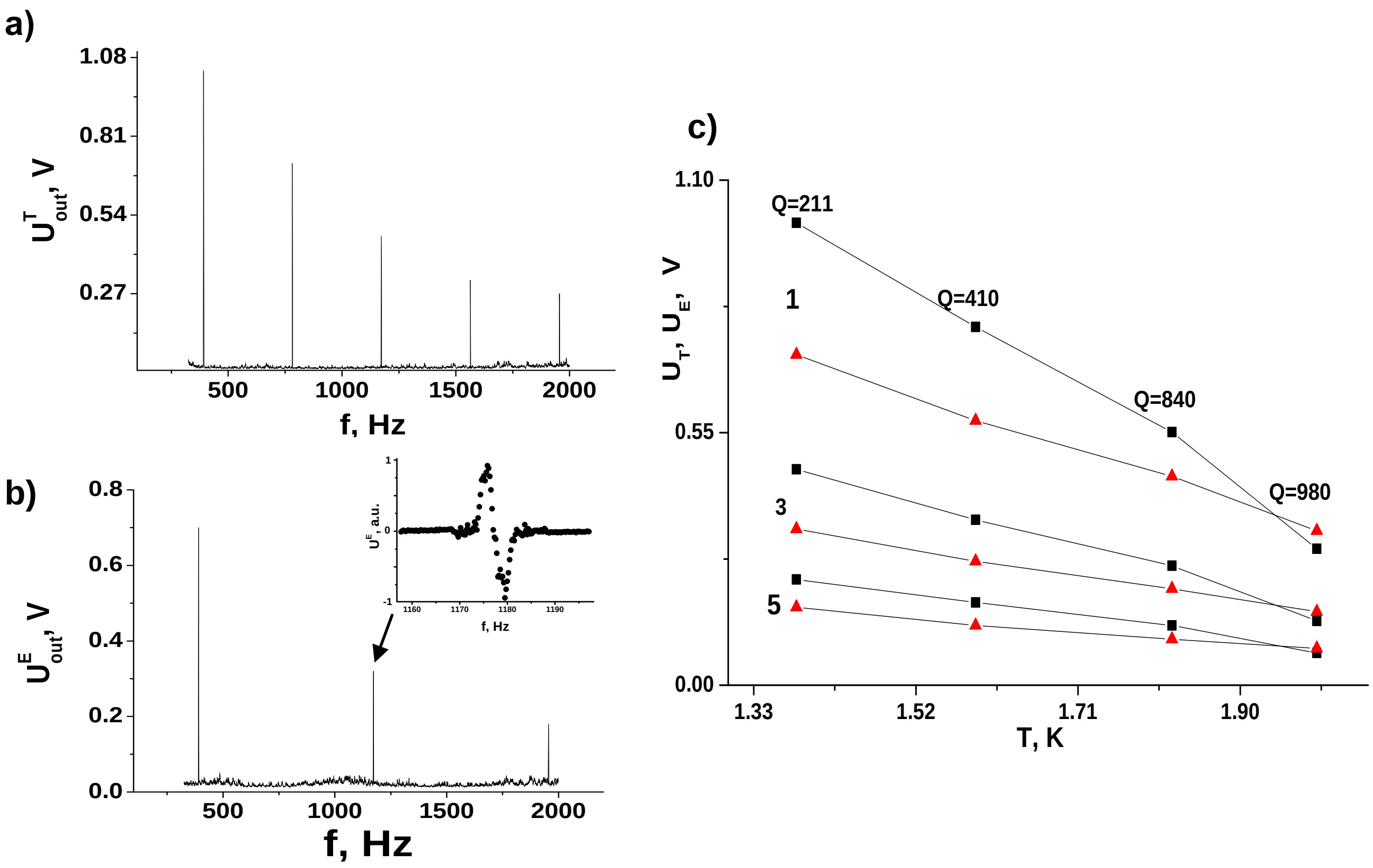

Fig.3 Amplitude-frequency characteristics of the temperature oscillations and the oscillations of electric potential. Specific power submitted to a heater is 3,22 mW/cm$^2$.The temperature is 1.39 K. The potential difference is measured between an electrode 9 and a cover of a heater. On an insert the derivative of the third mode is presented.

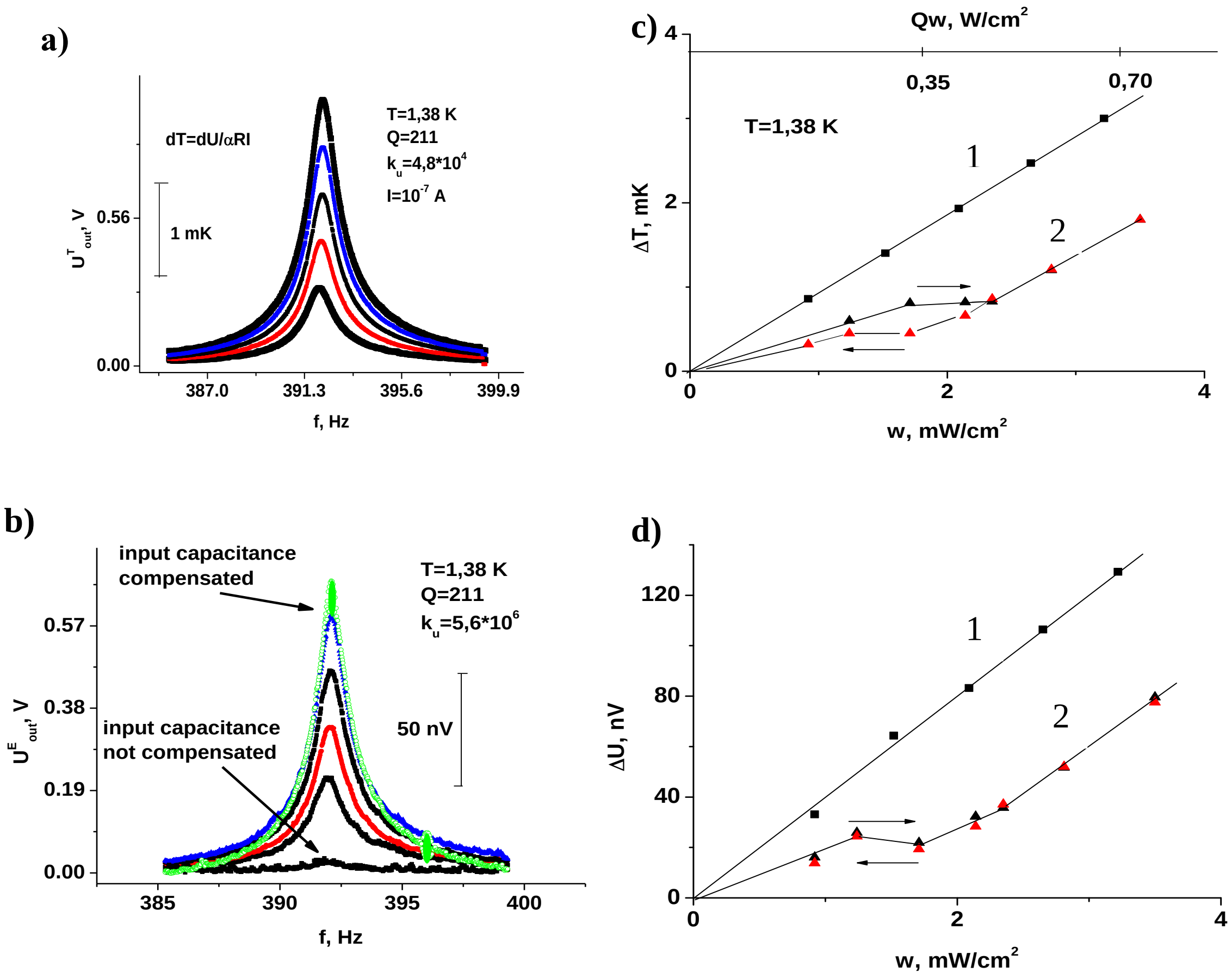

Fig.4 Influence of the capacity submitted to a heater, on amplitude of oscillations of temperatures (a) and the electric response (b). At the left: amplitude-frequency characteristics of the second sound resonator are presented (temperature 1,42 K, the capacities are 3,22, 2,65, 2,09, 1,52, 0,92 mW/cм$^2$).

The values of $U^E_{out}$(fig.b) are presented with taking into account the compensation of the entrance capacity of a preamplifier with the exception of the bottom curve when the capacity $C_{cal}$≈10 pF(Fig.1) was connected.

On the right: the capacity dependences of ΔT and ΔU for the first mode are presented. Triangles show the ΔT and ΔU results for a case, when the paper spacer 13 ( Fig. 1) was replaced by a macroscopic slit.

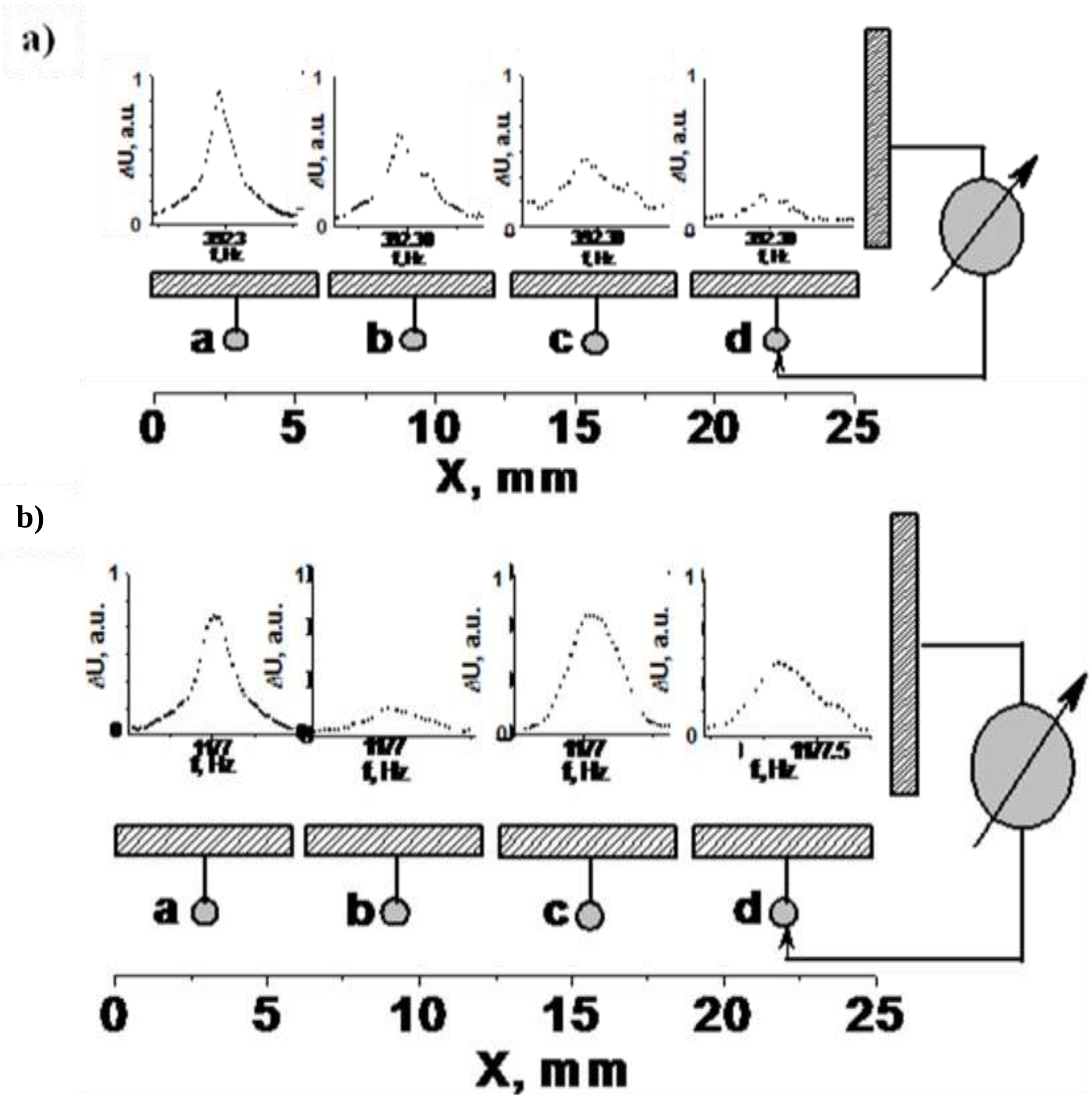

Fig. 5. Resonances of electric potential ΔU (f) between electrodes a, b, c, d and 9 for conditions, when a length of the resonator is equal to λ/2 (the top figure) and 3λ/2 (the bottom figure).

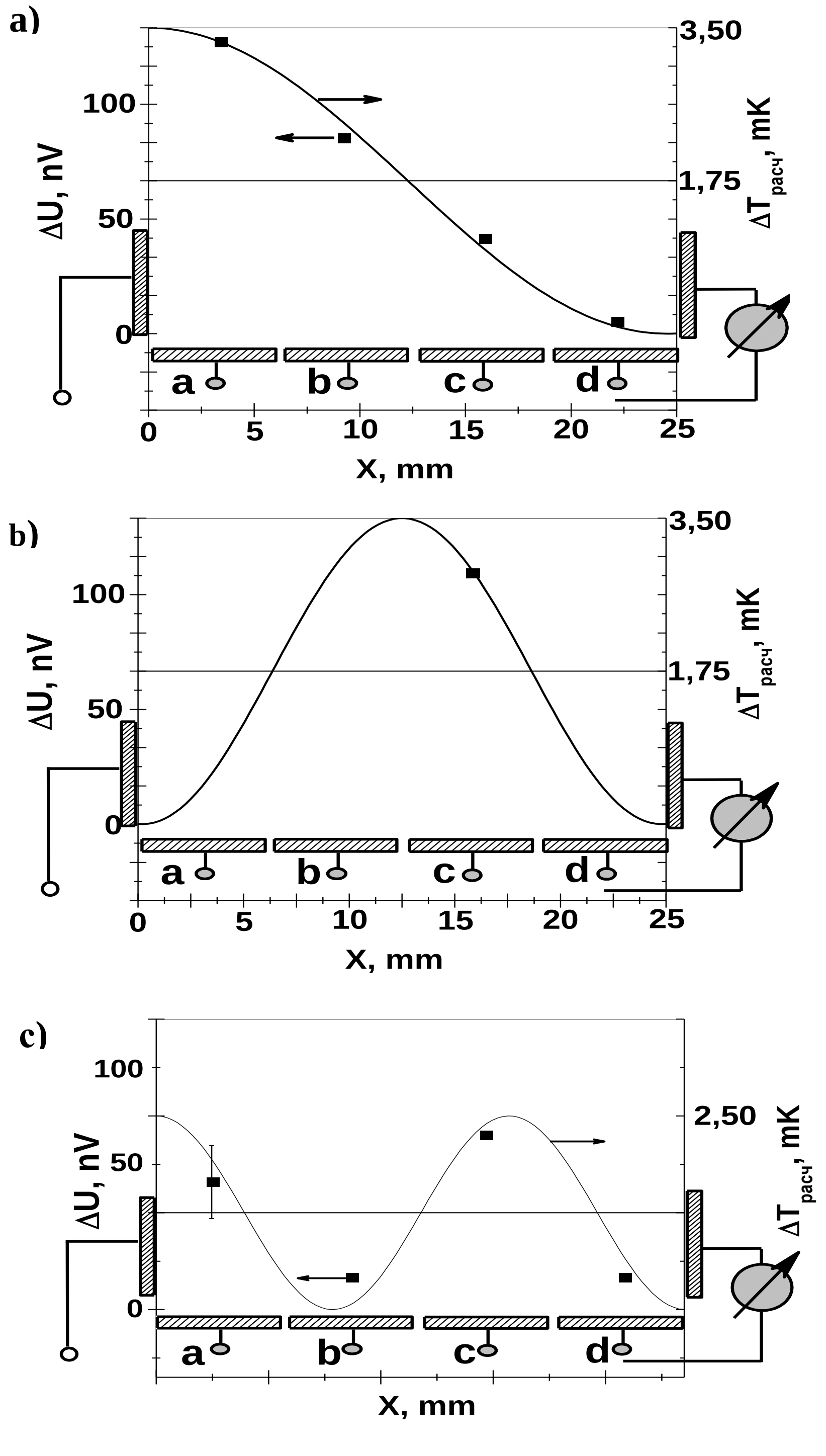

Fig. 6. Potential difference between electrodes a, b, c, d and 9. Square points are experimental results (the scale is shown by an arrow to the left) on resonant frequency as a result of averaging on 100 points. A solid line is calculation of spatial distribution of temperature in a second sound wave (the scale is shown by an arrow to the right).
a) – f=392 Hz, b) -f=784 Hz, c) – f=1176 Hz.

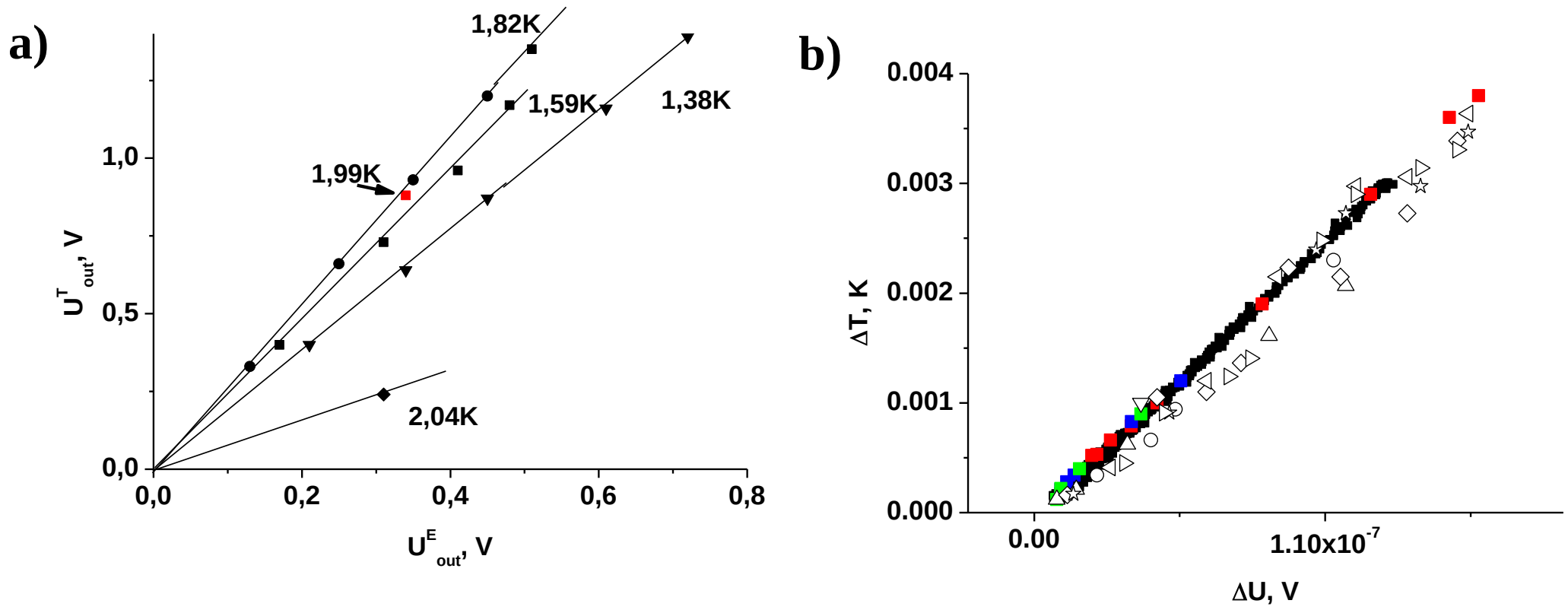

Fig. 7.a – Dependence between the voltages on the bolometer and electrode; b – dependences between the amplitude of temperature oscillations and, and amplitude of electric potential oscillations on the electrode 9.

■ – data of measurements between a cover of a heater and 9 given figures 3, ☆ – T=1,98 K, the open points - data of measurements between electrodes a, b, c, d and 9, red, dark blue and green points – the given works [1.] ΔTi against electrodes a b c d paid off, instead of was measured.